\documentclass{article}
\usepackage[margin = 1 in ]{geometry}
\usepackage{graphicx}
\usepackage{amsfonts}
\usepackage{amsmath}
\usepackage{amssymb}
\usepackage{dsfont}
\usepackage{xcolor}
\usepackage{hyperref}
\usepackage{bm}
\usepackage{ragged2e}

\usepackage[style=authoryear,natbib=true,sorting=nyt,backend=bibtex]{biblatex}
\newcommand\noneq{\text{neq}}

\newcommand\diff{\text{d}}

\newcommand\fL{\mathcal{L}}
\newcommand{\fH}{\mathcal{H}}

\newcommand\Indic{\mathds{1}}

\newcommand\avg[1]{\left\langle #1 \right\rangle}
\newcommand\avgph[1]{\avg{#1}_{\hat{p}}}
\newcommand{\pd}{\partial}

\begin{document}
{\centering \textbf{\LARGE On a structure preserving closure of Langevin dynamics}\\
\vspace{6 pt}
Travis Leadbetter${}^1$, Prashant K. Purohit${}^2$, and Celia Reina${}^2$\footnote[1]{Corresponding author: Celia Reina, Department of Mechanical Engineering and Applied Mechanics, University of Pennsylvania, 220 South 33rd Street, 229 Towne Building, Philadelphia PA, 19104, US.}\\
\vspace{12 pt}
{\small 
${}^1$ Graduate Group in Applied Mathematics and Computational Science, University of Pennsylvania, Philadelphia PA, 19104, US.\\
${}^2$ Department of Mechanical Engineering and Applied Mechanics, Univerisity of Pennsylvania, Philadelphia PA, 19104, US.\\

\vspace{12 pt}

ORCID: T.L.\ 0000-0001-7389-6248, P.K.P.\ 0000-0002-1939-907X, C.R.\ 0000-0001-9302-3401.\\

\vspace{12 pt} 

Word count: 4949, number of tables: 1, number of figures: 3 (including one video in Appendix E), this article has a supplementary video (mp4) to be included in section E of the appendix.\\
}
\vspace{24pt}
\textbf{\Large Abstract}\\
\justify{Given a particle system obeying overdamped Langevin dynamics, we demonstrate that it is always possible to construct a thermodynamically consistent macroscopic model which obeys a gradient flow with respect to its non-equilibrium free energy. To do so, we significantly extend the recent Stochastic Thermodynamics with Internal Variables (STIV) framework, a method for producing macroscopic thermodynamic models far-from-equilibrium from the underlying mesoscopic dynamics and an approximate probability density of states parameterized with so-called internal variables. Though originally explored for Gaussian probability distributions, we here allow for an arbitrary choice of the approximate probability density while retaining a gradient flow dynamics. This greatly extends its range of applicability and automatically ensures consistency with the second law of thermodynamics, without the need for secondary verification. We demonstrate numerical convergence, in the limit of increasing internal variables, to the true probability density of states for both a multi-modal relaxation problem, a protein diffusing on a strand of DNA, and for an externally driven particle in a periodic landscape. Finally, we provide a reformulation of STIV with the quasi-equilibrium approximations in terms of the averages of observables of the mesostate, and show that these, too, obey a gradient flow.}

\vspace{24 pt}
\begin{center}
Stochastic Thermodynamics, Thermodynamics with Internal Variables, Gradient Flow
\end{center}
} 
\clearpage
\justify{
\section{Introduction}
The field of non-equilibrium thermodynamics boldly seeks to describe the macroscopic behavior of a wide range of physical systems. Using various formulations, such as irreversible thermodynamics \citep{jou1996,lebon2008understanding}, rational thermodynamics \citep{coleman1964,truesdell1984historical}, thermodynamics with internal variables \citep{maugin1994}, or the General Equation for Non-Equilibrium Reversible-Irreversible Coupling (GENERIC) formalism \citep{grmela1997dynamics,mielke2011formulation,pavelka2018multiscale}, complex phenomena like creep \citep{garofalo1963empirical}, damage \citep{chaboche1981continuous,lemaitre1985continuous,bazant1986mechanics}, thermoplatiscity \citep{mielke2011formulation}, and fluid rheology \citep{ottinger2005beyond}, to name a few, have been modeled effectively and systematically. However, central to our understanding, and at the heart of Hilbert's sixth problem, is the ability to formally and practically connect these macroscopic models to more fundamental microscopic physics. Perhaps the two most important contributions on this front are the Mori-Zwanzig formalism and other projection operator approaches which seek to project out noisy or irrelevant aspects of a Hamiltonian system, leaving only quantities which vary over long time scales, and the utilization of large deviation principles to coarse-grain microscopic stochastic processes \citep{zwanzig2001nonequilibrium,ottinger1998general,montefusco2021framework,mielke2024deriving}. Although these two theories provide the cornerstone of our theoretical understanding, significant challenges limit their utility in producing usable models. 
\par 
In a previous work \citep{leadbetter2023}, the authors leveraged the theory of stochastic thermodynamics and a variational principle to derive thermodynamically consistent macroscopic models directly from microscopic particle systems obeying an overdamped Langevin dynamics of the form
\begin{equation}\label{eqn:sde}
    \diff x_t = -\frac{1}{\eta}\pd_x e \,  \diff t + \sqrt{\frac{2}{\eta\beta}}\,\diff w_t,
\end{equation}
where the interaction energy $e(x,\lambda_t)$ is potentially dependent on an external driving protocol $\lambda_t$, $\eta$ and $\beta$ are the viscosity and inverse temperature respectively (both assumed to be constant), and $w_t$ is a standard Brownian motion. The resulting macroscopic models exhibited the internal variable structure for an arbitrary choice of approximation to the probability density of states, and hence the framework was named Stochastic Thermodynamics with Internal Variables (STIV).  For the particular choice of a Gaussian approximation to the probability density of states, the dynamics of the internal variables obeyed a gradient flow with respect to the non-equilibrium free energy, $\hat{A}^\noneq$, thereby guaranteeing a non-negative rate of total entropy production.
In this work, we show that the Gaussian approximation is not an isolated example, but that the gradient flow structure can be obtained for any arbitrary choice of approximation to the probability density of states. This is fitting, as it is well known that the Fokker-Planck equation associated with Langevin dynamics can be formulated as a gradient flow of the free energy $A^\text{neq}=\int \left(e+\beta^{-1} \log p \right) p \, dx $ with respect to the Wasserstein-2 metric \citep{jordan1998variational}
\begin{equation}
    \pd_t p = \frac{1}{\eta} \partial_{x_i} \left(p\ \pd_{x_i} \frac{\delta A^{\text{neq}} }{\delta p}  \right)
\end{equation}
Thus, this enhanced STIV method can preserve this fundamental structure, eliminating the need to verify consistency with the second law of thermodynamics, for example by using the Coleman-Noll procedure \citep{coleman1963thermodynamics}.
\section{Derivation} 
The dynamical equations for the STIV framework are derived from the variational method of Eyink \citep{eyink1996}, which provides a method for approximating the solution to differential equations arising in non-equilibrium statistical mechanics. Starting from the Fokker-Planck equation associated with the microscopic stochastic differential equation in Eq. \ref{eqn:sde}\footnote[1]{We shall use the Einstein summation convention, summing all repeated indices, but without regard for raised and lowered indices.}
 \begin{align*}
     \pd_t p(x,t) &= \fL p(x,t) = \pd_{x_i} \left( \frac1{\eta}p(x,t)\pd_{x_i}e + \frac1{\eta\beta}\pd_{x_i} p(x,t) \right)
 \end{align*} 
 one can recast the problem in a variational form by introducing a test function $\psi$, and the action 
 \begin{equation*}
     \Gamma[\psi,p] = \int_0^\infty \int_X \psi\left(\pd_t - \fL\right)p \,\diff x\diff t.
 \end{equation*}
Stationary points $(\psi^*,p^*)$ of this action are solutions to the adjoint and standard forms of the original Fokker-Planck equation respectively, and solve an infinite dimensional Hamilton equation 
\begin{align*}
    \pd_t p &= \frac{\delta}{\delta \psi} \fH[\psi,p] \\
    \pd_t \psi &= -\frac{\delta}{\delta p} \fH[\psi,p]
\end{align*}
where $\fH = \int_X \psi \fL p \,\diff x$ is the Hamiltonian. The adjoint equation is trivially solvable with solution $\psi^* \equiv 1$. The equation for $p$ is intractable in general, so to find an approximate solution one proposes a parameterized approximate probability density of states, $\hat{p}(x,\alpha)$, and a parameterized test function, $\hat{\psi}(x,\alpha)$. Writing $\hat{u} = \log(\hat{p})$ and $\avg{f}_{\hat{p}} =\int f(x)\hat{p}(x,\alpha)\diff x$, the dynamical equations for the parameters $\alpha$ reduce to 
\begin{equation*}
\avg{[\pd\hat{\psi},\pd\hat{u}]_{ij}}_{\hat{p}} \dot{\alpha}_j = \pd_{\alpha_i}\avg{\fL^\dagger \hat{\psi}}_{\hat{p}}
\end{equation*}
where $[\pd\hat{\psi},\pd\hat{u}]_{ij} = \pd_{\alpha_i}\hat{\psi}\pd_{\alpha_j}\hat{u} - \pd_{\alpha_j}\hat{\psi}\pd_{\alpha_i}\hat{u}$ and $\fL^\dagger$ is the adjoint of $\fL$. The matrix $\avgph{[\pd\hat{\psi},\pd\hat{u}]_{ij}}$ is anti-symmetric and obeys the Jacobi identity, so these dynamics retain the Hamiltonian form. 
\par
In the original framework \citep{leadbetter2023}, the parameters were split into two sets: the variables parameterizing $p$ denoted (with a slight abuse of notation) $\alpha$ and referred to as internal variables, and dummy variables $\gamma$ of the same dimension as $\alpha$. The approximate density is assumed to only depend on $\alpha$ while $\hat{\psi}$ was defined as $\hat{\psi} = 1 + \gamma_i\, \pd_{\alpha_i} \hat{s}$, where $\hat{s} = -k_B\hat{u} = - k_B \log(\hat{p})$ is the approximate microscopic entropy, and $k_B$ is the Boltzmann constant. With this particular choice, the dynamical equations reduce to 
\begin{equation*}
\avgph{\pd_{\alpha_i}\hat{s}\, \pd_{\alpha_j}\hat{s}} \dot{\alpha}_j = -k_B \avgph{\fL^\dagger \pd_{\alpha_i}\hat{s}},\qquad\gamma(t) \equiv 0.
\end{equation*}
This ``linear ansatz'' using the $\alpha$-gradient of the microscopic entropy in the test function $\hat{\psi}$ produces a specific case of the moment-closure equations called ``entropic matching'' \cite{bronstein2018variational}, and ultimately leads to irreversible dynamics for $\alpha$. 
\par
Once the dynamics for the internal variables are obtained, one can approximate macroscopic thermodynamic quantities defined via stochastic thermodynamics by substituting the approximate probability density of states for the true solution. For example, the macroscopic non-equilibrium free energy defined as $A^\noneq = \avg{e}_p - T\avg{s}_p$ becomes $\hat{A}^\noneq = \avgph{e} - T\avgph{\hat{s}}$, from which one can approximate both the external force $F^{\text{ex}} = \avg{\pd_\lambda e}_p$ by $\hat{F}^{\text{ex}} = \pd_\lambda \hat{A}^\noneq$ and the rate of total entropy production $\frac{\diff}{\diff t} S^{\text{tot}}$ by $\frac{\diff}{\diff t} \hat{S}^{\text{tot}} = \frac{1}{T}\left( \hat{F}^{\text{ex}} \dot{\lambda} - \frac{d}{dt}\hat{A}^\noneq \right)=-\frac1{T}\pd_{\alpha_i} \hat{A}^\noneq \dot{\alpha}_i$. 
\par
Previously, the STIV framework was effectively used to study the dynamics and thermodynamics of phase front propagation in an elastic medium, where one could derive both the continuum limit \citep{leadbetter2023}, and an equivalent phase field model directly from microscopic physics \citep{leadbetter2024statistical}. This previous implementation, however, was limited to a multivariate Gaussian approximation to the probability density of states $\hat{p}$ as this was the only form in which the gradient flow structure, and hence non-decreasing total entropy production, could be ensured at the time. 
The key findings of this work are a number of options for preserving the gradient flow structure of the underlying Fokker-Planck equation which are flexible enough to approximate any probability density to arbitrary accuracy in the limit of increasing internal variables. The first method, which we call the ``operator method'', utilizes an alternative form for the approximate test function that allows for a completely unconstrained choice of approximate probability density of states. The second uses the original approximate test functions and entropic matching moment-closure equations, but requires one to restrict the approximate probability density of states using a ``quasi-equilibrium'' distributions \citep{turkington2013}. The final method forgoes the need to include the Boltzmann factor $\exp(-\beta e(x,\lambda_t))$ in the approximate probability density of states, which could reduce computational costs in some cases, but the internal variable dynamics given by entropic matching  only approximate a gradient flow. We label this method as ``exponential family'' throughout. A detailed analysis of each of these three methods is provided next, and the key elements and results for each of them are summarized in Table \ref{table:results} as reference. 
{\renewcommand{\arraystretch}{1.2}
\begin{table*}
\footnotesize
\begin{center}
    \caption{Key features of the closure methods$^*$}
    \label{table:results}
    \begin{tabular}{|c|c c c|} 
    \hline 
        Method Name & Operator method & Quasi-equilibrium & Exponential family \\
     \hline   
        Probability density of states $\hat{p}$ & Arbitrary & $\exp(-\beta e + \alpha \phi - f)$ & $\exp(\alpha \phi - f)$ \\
    \hline
        Test functions $\hat{\psi}$ & $1 + \gamma (\fL_{\hat{u}}^\dagger)^{-1} \pd_\alpha \hat{u}$ & $1 + \gamma \pd_\alpha \hat{u}$ & $1 + \gamma \pd_\alpha \hat{u}$ \\
    \hline
        Dynamical equations & $\mathbb{F}\dot{\alpha} = -\frac1{\eta}\pd_{\alpha}\hat{A}^\noneq$ & $\mathbb{C}\dot{\alpha} = -\frac1{\eta}\mathbb{M}\mathbb{C}^{-1}\pd_{\alpha}\hat{A}^{\noneq}$ & $\mathbb{C}\dot{\alpha} = -\frac1{\eta}\mathbb{M}\mathbb{C}^{-1}\pd_{\alpha}\hat{A}^{\noneq}{}^\S$ \\
    \hline 
        Dynamics with external driving & $\mathbb{F}\dot{\alpha} = -\frac1{\eta}\pd_{\alpha}\hat{A}^\noneq$ & $\mathbb{C}\dot{\alpha} + \avgph{\bar{\phi}\pd_\lambda \hat{u}}\!\!\!\dot{\lambda} 
        = -\frac1{\eta}\mathbb{M}\mathbb{C}^{-1}\pd_\alpha \hat{A}^{\noneq}$ & $\mathbb{C}\dot{\alpha} = -\frac1{\eta}\mathbb{M}\mathbb{C}^{-1}\pd_\alpha \hat{A}^{\noneq}$ \\
    \hline
        Dual dynamics & Not studied here & $\frac{\diff}{\diff t} \hat{\Phi} = -\frac1{\eta}\mathbb{M}\pd_{\Phi} \hat{A}^{\noneq}{}^\ddag$ & $\frac{\diff}{\diff t} \hat{\Phi} = -\frac1{\eta}\mathbb{M}\pd_{\Phi} \hat{A}^{\noneq}$ \\
    \hline
        Entropy production $T \frac{\diff}{\diff t} \hat{S}^\text{tot}$ & $\eta  \mathbb{F}_{ij}\dot{\alpha}_i \dot{\alpha}_j$ & $ \frac1{\eta\beta^2}\mathbb{M}_{ij}\alpha_i\alpha_j$& $\eta \mathbb{M}_{ij}^{-1}\frac{\diff}{\diff t}\hat{\Phi}_i\frac{\diff}{\diff t}\hat{\Phi}_j$ \\ 
    \hline
    \end{tabular}
    \vspace{6 pt} \\ 
    {}$^*$ Definition of terms for Table \ref{table:results}:\\ $\mathbb{C}_{ij} = \avgph{\pd_{\alpha_i}\hat{u}\pd_{\alpha_j}\hat{u}} = \text{Cov}_{\hat{p}}(\phi_i,\phi_j)$,  $\mathbb{M}_{ij} = \avgph{\pd_{x_k}\pd_{\alpha_i}\hat{u}\pd_{x_k}\pd_{\alpha_j}\hat{u}} = \avgph{\pd_{x_k}\phi_i\pd_{x_k}\phi_j}$,\\
    $\fL_{\hat{u}}^\dagger g = \pd_{x_i} \hat{u} \pd_{x_i} g + \pd_{x_i} \pd_{x_i} g$ , $\mathbb{F}_{ij} = \avgph{\pd_{\alpha_i}\hat{u} \,(\fL_{\hat{u}}^\dagger)^{-1}\pd_{\alpha_j}\hat{u}}$, $\bar{\phi} =\phi - \hat{\Phi}$, $\hat{\Phi}=\avgph{\phi}$. \\
    ${}^\ddag$ This equation holds with and without external driving.\\
    $^\S$ These dynamics are not given by moment closure, but approximate it for large number of observables.
\end{center}
\end{table*}
}

\subsection{The operator method}
We now motivate an alternative choice for the approximate test function $\hat{\psi}$ to show how the STIV method can produce gradient flow equations for an arbitrary approximate probability density of states. Again, we fix an approximate density $\hat{p}(x,\alpha)$, and let $\hat{u} = \log(\hat{p})$. We then assume that $\chi(x,\alpha)$ is a vector valued function with the same dimension as $\alpha$, and define $\hat{\psi}(x,\alpha,\gamma) = 1 + \gamma_i\, \chi_i(x,\alpha)$. Following the same steps as above, the dynamical equations induced by the variational method become 
\begin{equation*}
\avgph{\chi_i\, \pd_{\alpha_j}\hat{u}} \dot{\alpha}_j = \avgph{\fL^\dagger \chi_i},\qquad \gamma(t) = 0.
\end{equation*}
For convenience, we define $\mathbb{F}_{ij} \equiv \avgph{\chi_i\, \pd_{\alpha_j}\hat{u}}$ and $\mathbb{C}_{ij} \equiv \avgph{\pd_{\alpha_i}\hat{u}\, \pd_{\alpha_j}\hat{u}}$.  
\par
We have not specified new test functions yet, but to do so we examine the approximate rate of total entropy production, $T\dot{\hat{S}}^\text{tot} = -\dot{\alpha}_i\, \pd_{\alpha_i} \hat{A}^\noneq$. We recall that $\hat{A}^\noneq = \hat{E} - T\hat{S} = \avgph{e} + \avgph{\hat{u}/\beta}$, and since $0 = \pd_{\alpha_i}\avgph{1} = \avgph{\pd_{\alpha_i}\hat{u}}$ and $e$ does not depend on $\alpha$, we have $\pd_{\alpha_i}\hat{A}^\noneq = \avgph{(e + \hat{u}/\beta)\pd_{\alpha_i}\hat{u}}$. On the other hand, repeated integration by parts reveals that 
\begin{align*}
    \avgph{\fL^\dagger \chi_i} &= \avgph{\left(-\frac1{\eta}\pd_{x_k} e \pd_{x_k} \chi_i + d \pd_{x_l}\pd_{x_l} \chi_i \right)} \\
    &=-\frac1{\eta}\avgph{\pd_{x_k}\left(e + \hat{u}/\beta\right)\pd_{x_k}\chi_i } \\
    &= \frac1{\eta}\avgph{(e + \hat{u}/\beta)\left(\pd_{x_k}\hat{u}\pd_{x_k}\chi_i + \pd_{x_k}\pd_{x_k}\chi_i \right)}\\
    &\equiv\frac1{\eta} \avgph{(e + \hat{u}/\beta)\fL_{\hat{u}}^\dagger \chi_i}.
\end{align*}
This new operator $\fL_{\hat{u}}^\dagger g = \pd_{x_k}\hat{u}\pd_{x_k}g + \pd_{x_l}\pd_{x_l}g$ is also the adjoint of a Fokker Planck operator which only depends on $\hat{u}$.  The final line resembles the equation for $\pd_{\alpha_i}\hat{A}^\noneq$, and suggests that one should choose test functions which obey
\begin{equation}\label{eqn:TestFunctionPDE}
\fL_{\hat{u}}^\dagger \chi_i = -\pd_{\alpha_i}\hat{u}.
\end{equation}
Assuming one can solve this differential equation, the dynamical equations become a gradient flow 
\begin{equation}\label{eqn:Dynamic1}
\mathbb{F}_{ij}\dot{\alpha}_j = -\frac1{\eta}\pd_{\alpha_i}\hat{A}^\noneq, 
\end{equation}
since $\mathbb{F}_{ij}$ is positive semi-definite as 
\begin{align*}
\mathbb{F}_{ij} &= \avgph{\chi_i\pd_{\alpha_j}\hat{u}} \\
 &= -\avgph{\chi_i \fL_{\hat{u}}^\dagger\chi_j } \\
 &= -\int \chi_i \pd_{x_k}\left(\hat{p}\pd_{x_k}\chi_j\right)\diff x \\
 &= \avgph{\pd_{x_k}\chi_i \pd_{x_k}\chi_j }.
\end{align*}
Between the second and third lines, we have made use of the fact that  $\hat{p}\fL_{\hat{u}}^\dagger \chi_j = \pd_{x_k}\left(\hat{p} \pd_{x_k} \chi_j\right)$.
With the gradient flow structure comes the guarantee of non-negative rate of total entropy production
\begin{equation*}
T\dot{\hat{S}}^{tot} = \eta\mathbb{F}_{ij}\dot{\alpha}_i\dot{\alpha}_j \geq 0.
\end{equation*}
\par
This strategy allows for a completely general choice of approximate probability density of states $\hat{p}(x,\alpha)$. Solving Eq. \ref{eqn:TestFunctionPDE}, though, is challenging in general. However, following a well known trick \citep{risken1996fokkerplanck}, one can define a ``Hamiltonian'' conjugation of $\fL_{\hat{u}}^\dagger$ by 
\begin{equation*}
-\fH_{\hat{u}}g = \sqrt{\hat{p}}\ \fL_{\hat{u}}^\dagger \left(\frac{g}{\sqrt{\hat{p}}}\right) = \pd_x^2 g - v(x,\alpha)g
\end{equation*}
where $v = \pd_x^2 \hat{u}/2 + |\pd_x \hat{u}|^2/4$. Unlike $\fL_{\hat{u}}^\dagger$, $\fH_{\hat{u}}$ is self-adjoint with respect to the Lebesgue measure. Note, also that $\fH_{\hat{u}}$ defines a Schr\"{o}dinger operator, and it is immediately clear that if $(\lambda^*,g^*)$ is an eigenvalue-eigenvector pair for $\fH_{\hat{u}}$, then $(-\lambda^*,g^*/\sqrt{\hat{p}})$ is an eigenvalue-eigenvector pair for $\fL_{\hat{u}}^\dagger$. Hence, one can take advantage of existing eigenvalue solvers for quantum systems in order to compute the necessary test functions.  

\subsection{Quasi-equilibrium distribution}
The primary advantage of the previous method for producing a gradient flow dynamics is that it allows for an arbitrary choice of approximate probability density of states. The disadvantage is that the test functions must be computed as solutions to the differential equation in Eq.~\ref{eqn:TestFunctionPDE}. Alternatively, we show now that if one restricts the approximate probability density of states to a specific class known as quasi-equilibrium densities (a specific form of exponential family) \citep{turkington2013}, then one may use the original test functions $\chi_i = \pd_{\alpha_i}\hat{u}$ and still achieve a gradient flow (here, we have replaced $\hat{s}$ by $\hat{u}$ for notational simplicity). This is only a minor restriction as any continuous probability distribution can be approximated by an exponential family or quasi-equilibrium distribution to arbitrary accuracy (see Appendix C for one such proof).  
\par 
Assume that the approximate probability density of states takes a quasi-equilibrium form. That is 
\begin{equation}\label{eqn:QuasiEQ}
\hat{p}(x,\alpha,\beta) = \exp(-\beta e(x) + \alpha_i \phi_i(x) - f(\alpha,\beta) ),
\end{equation}
for some set of ``observables'' $\phi_i(x)$ and internal variables $\alpha_i(t) $. Here, $\beta$ (taken as a constant) and the internal variables $\alpha$, form the natural parameters of the exponential family, while the energy $e$ and observables $\phi$ form the ``sufficient statistics'' \citep{lehmann2006theory}. It will be helpful to define $\hat{\Phi}_i$ as the $\hat{p}$ averages of the observables, i.e., $\hat{\Phi}_i \equiv \avgph{\phi_i}$. 
\par 
To see the gradient flow structure, note first that in this case 
\begin{equation*}
\avgph{\chi_i\pd_{\alpha_j}\hat{u}}  = \avgph{(\phi_i - \hat{\Phi}_i)(\phi_j - \hat{\Phi}_j)} = \mathbb{C}_{ij}.
\end{equation*}
Second, since $e + \hat{u}/\beta = (\alpha_i \phi_i -f(\alpha,\beta))/\beta$, we have 
\begin{equation*}
\pd_{\alpha_i}\hat{A}^{neq} = \avgph{(e + \hat{u}/\beta)\pd_{\alpha_i}\hat{u}} = \frac{\mathbb{C}_{ij}\alpha_j}{\beta}.
\end{equation*}
Finally, again by integrating by parts 
\begin{align*}
\avgph{\fL^\dagger \chi_i} &= -\frac1{\eta}\avgph{\pd_{x_k}(e + \hat{u}/\beta)\pd_{x_k}\chi_i} \\
&= -\frac{\avgph{\pd_{x_k} \phi_i \pd_{x_k} \phi_j}\alpha_j}{\eta\beta} \\
&\equiv - \frac1{\eta\beta}\mathbb{M}_{ij}\alpha_j.
\end{align*}
As with $\mathbb{F}$ in the earlier discussion, $\mathbb{M}_{ij} = \avgph{\pd_{x_k} \phi_i \pd_{x_k} \phi_j}$ is positive semi-definite. Putting the pieces together, we get 
\begin{equation}\label{eqn:dynQE}
    \mathbb{C}_{ij}\dot{\alpha}_j = -\frac1{\eta\beta}\mathbb{M}_{ij}\alpha_j
\end{equation}
or, by inserting the identity as $\delta_{jk} = \mathbb{C}^{-1}_{jl}\mathbb{C}_{lk}$ between $\mathbb{M}$ and $\alpha$ (we assume $\mathbb{C}$ is invertible),
\begin{equation}\label{eqn:gradFlow}
    \dot{\alpha}_i = -\frac1{\eta}\mathbb{C}^{-1}_{ij}\mathbb{M}_{jk}\mathbb{C}^{-1}_{kl}\pd_{\alpha_l}\hat{A}^\noneq.
\end{equation}
Since $\mathbb{M}$ is positive semi-definite, and $\mathbb{C}$ is symmetric, $\mathbb{C}^{-1}\mathbb{M}\mathbb{C}^{-1}$ is positive semi-definite and the dynamics obey a gradient flow. 

\subsection{Exponential family}
Using a quasi-equilibrium distribution which includes the total energy leads to gradient flow dynamics exactly. However, one may not always wish to include the energy in the approximate density of states, for example, if it would make computing the normalizing function more challenging. If one removes the energy and uses an exponential family of the form
\begin{equation}\label{eqn:ExpFamily}
    \hat{p}(x,\alpha) = \exp(\alpha_i \phi_i - f(\alpha))
\end{equation}
the dynamics are not always a gradient flow. Specifically, they become
\begin{align}\label{eqn:dynNoE}
    \mathbb{C}\dot{\alpha} &= -\frac1{\eta}\mathbb{M} \mathbb{C}^{-1} \pd_{\alpha} \hat{A}^\noneq \\
    &\quad - \frac1{\eta}\left(\avgph{\pd_{x_k} e \pd_{x_k} \phi} - \mathbb{M}\mathbb{C}^{-1}\avgph{e(\phi - \hat{\Phi})}\right). \nonumber
\end{align}
The second term on the right hand side only depends on the (non-constant) part of the energy which is $\hat{p}$-orthogonal to the span of the $\phi$. Mathematically, if we can write $e(x) = \gamma_j (\phi_j(x) - \hat{\Phi}_j) + e_0 + e^\perp(x)$ where $\avgph{e^\perp(\phi_j - \hat{\Phi}_j)} = 0$ for all $j$, then the second term becomes 
\begin{equation*}
    \left(\avgph{\pd_{x_k} e \pd_{x_k} \phi} - \mathbb{M}\mathbb{C}^{-1}\avgph{e(\phi - \hat{\Phi})}\right) = \avgph{\pd_{x_k} e^\perp \pd_{x_k} \phi}.
\end{equation*}
If the observables are chosen to produce a highly flexible approximation, for example by using machine learning techniques or a basis expansion, it is reasonable to assume that this term is negligible and therefore one can set it to zero, giving a gradient flow structure. In this case, $\pd_{\alpha}\hat{A}^\noneq = \mathbb{C}\alpha/\beta + \avgph{e(\phi - \hat{\Phi})}$, and again the dynamical equations are given by Eq.~\ref{eqn:gradFlow}. To minimize confusion, we shall refer to an approximate probability density of states of the form $\hat{p} = \exp(\alpha_i \phi_i - f(\alpha))$ which does not include the energy as an exponential family approximation. When $\hat{p} = \exp(-\beta e + \alpha_i \phi_i - f(\alpha,\beta))$ explicitly includes the energy, we shall refer to this as a quasi-equilibrium approximation (although it is also an exponential family). 
 
\subsection{Time dependent external driving}
Now we consider the case when the total energy is time dependent through some external driving protocol, $e = e(x,\lambda(t))$. In the operator formulation, with test functions $\fL_u \chi_i = \pd_{\alpha_i} \hat{u}$, nothing needs change to maintain the original STIV definitions. Since the approximate probability density of states can still be chosen to be independent of $\lambda$ (for fixed $\alpha$), the approximate external force remains $\hat{F}^{\text{ex}} = \pd_\lambda\hat{A}^\noneq$ \citep{leadbetter2023}. 
In the quasi-equilibrium formulation, $\hat{p}$ explicitly depends on $\lambda$ through $e$ and so it is no longer true that $\hat{F}^\text{ex} = \pd_\lambda \hat{A}^\noneq$. Moreover, when deriving the dynamical equations from the variational method \citep{eyink1996}, it was assumed that all of the time dependence of $\hat{p}$ was captured through the internal variables $\alpha$. When $\hat{p}$ takes the quasi-equilibrium form this is not the case since the energy is now time dependent. This is fixed through the proper modification of the dynamical equations given in Eq. \ref{eqn:dynQE} (which can be found in Appendix A).
The resulting dynamical equations are
\begin{equation}\label{eqn:dynFex}
    \mathbb{C}_{ij}\dot{\alpha}_j + \avgph{(\phi_i - \hat{\Phi}_i)\pd_\lambda \hat{u}}\dot{\lambda} = -\frac1{\eta\beta}\mathbb{M}_{ik}\alpha_k
\end{equation}
and the external force becomes
\begin{equation}\label{eqn:FEx}
    \hat{F}^\text{ex} = \pd_\lambda \left( \hat{A}^\noneq - \frac{ \alpha_i\hat{\Phi}_i}{\beta}\right) = -\frac1{\beta}\pd_\lambda  f(\alpha,\beta,\lambda).
\end{equation}
The equation for the rate of total entropy production is also impacted, but it still remains non-negative 
\begin{equation}\label{eqn:EPQEq}
    T\frac{\diff}{\diff t} \hat{S}^\text{tot} = \frac1{\eta\beta^2}\alpha_i \mathbb{M}_{ij}\alpha_j
\end{equation}
(see the Appendix B for the derivation of both the approximate external force and approximate rate of total entropy production). 
\par
The new dynamical equations appear to break the gradient flow structure. However, in the following section, we show that the gradient flow structure remains intact when the equations are recast in terms of the dynamics of the averages of observables $\hat{\Phi}$. 

\subsection{Dual formulation}
So far, the dynamics of the system have been tracked through the internal variables $\alpha$. However, in traditional internal variable formulations, the internal variables are generally assumed to be averages of microscopic substructures which are not captured in equilibrium. Thus, it seems more natural to consider the averages of the observables, that is $\hat{\Phi}_i = \avgph{\phi_i}$ as independent variables, and develop the system's dynamics in terms of their behavior. Fortunately, this is readily achievable through a simple change of variables. Assuming the map from the internal variables to the averages of the observables, $\alpha \rightarrow \hat{\Phi}$ is invertable, we can define the non-equilibrium free energy as a function of $\hat{\Phi}$. Using the chain rule, we find that $\pd_{\alpha}\hat{A}^\noneq = \mathbb{C}\pd_{\Phi}\hat{A}^\noneq$. Combining this with 
\begin{align*}
    \frac{\diff}{\diff t} \hat{\Phi}_i &= \avgph{\phi_i\pd_{\alpha_j}u}\dot{\alpha}_j  + \avgph{\phi_i\pd_\lambda \hat{u}}\dot{\lambda}\\
    &= \mathbb{C}_{ij}\dot{\alpha}_j +\avgph{(\phi_i - \hat{\Phi}_i)\pd_\lambda \hat{u}}\dot{\lambda},
\end{align*}
which reproduces the left hand side of Eq.~\ref{eqn:dynFex} (or Eq.~\ref{eqn:dynQE} and Eq.~\ref{eqn:dynNoE} when $\hat{p}$ doesn't depend on $\lambda$), gives the equivalent dynamics for $\hat{\Phi}$ 
\begin{equation*}
\frac{\diff}{\diff t} \hat{\Phi}_i = - \frac1{\eta}\mathbb{M}_{ij} \pd_{\Phi_j}\hat{A}^\noneq
\end{equation*}
which again have the gradient flow structure. This holds true for both the quasi-equilibrium and exponential family approximations. Moreover, when $\hat{p}$  is independent of $\lambda$ (either the exponential family approximation or the quasi-equilibrium approximation with no external driving), the form of the rate of total entropy production is invariant under the change of variables \citep{leadbetter2024statistical} 
\begin{align*}
    T\frac{\diff}{\diff t}\hat{S}^{\text{tot}} &= -\pd_{\alpha_i}\hat{A}^\noneq (\alpha_i,\beta)\dot{\alpha}_i \\
    &= - \pd_{\Phi_i} \hat{A}^\noneq(\hat{\Phi},\beta) \frac{\diff \hat{\Phi}_i}{\diff t} \\
    &= \eta \frac{\diff \hat{\Phi}_i }{\diff t} \mathbb{M}_{ij}^{-1} \frac{\diff \hat{\Phi}_j}{\diff t}.
\end{align*}
In this form, it becomes clear that $\mathbb{M}_{ij} = \avgph{\pd_x \phi_i \pd_x \phi_j}$ is an approximation to the true average dissipation matrix of the observables $\phi_i(x_t)$ along fluctuating trajectories. Mathematically, since the governing Langevin equation is given by 
\begin{equation*}
    \diff x_t = -\pd_x e /\eta\,  \diff t + \sqrt{2/\eta\beta}\,\diff w_t,
\end{equation*}
we know by It\^{o}'s formula that
\begin{equation*}
    \diff \phi_i(x_t) = \left(- \frac1{\eta}\pd_x e \pd_x\phi_i + \frac{1}{\eta\beta}\Delta_x \phi_i \right)\,  \diff t + \sqrt{\frac{2}{\eta\beta}}\pd_x \phi_i \, \diff w_t
\end{equation*}
and hence the diffusion matrix of $\phi(x_t)$ is $\mathcal{M}_{ij}(x_t) \propto \pd_x \phi_i \pd_x \phi_j$. When $\hat{p}$ depends on $\lambda$, the approximate rate of total entropy production cannot be written as a quadratic form in $\frac{\diff}{\diff t} \hat{\Phi}$ but is still given by Eq.\ \ref{eqn:EPQEq}. 
\par
It is also interesting to note that in the case of the quasi-equilibrium approximation, the non-equilibrium free energy is the convex dual of the normalizing function
\begin{equation*}
f(\alpha,\beta,\lambda) = \log(\int \exp(-\beta e + \alpha \phi)\diff x).
\end{equation*}
$f$ is convex in $\alpha$ since $\pd_{\alpha_i}\pd_{\alpha_j} f = \avgph{(\phi_i - \hat{\Phi}_i)(\phi_j - \hat{\Phi}_j)} = \mathbb{C}_{ij}$ is positive semi-definite and we know
\begin{equation*}
 \hat{\Phi}_i = \avgph{\phi_i} = \pd_{\alpha_i}f.
\end{equation*}
Using the Legendre transform, we can define the convex dual to $f$ as $ f^*(\Phi,\beta,\lambda) = \sup_{\alpha}\big(\alpha \Phi - f(\alpha,\beta,\lambda) )$ so that 
\begin{equation*}
\alpha_i = \pd_{\Phi_i} f^*(\Phi = \hat{\Phi},\beta,\lambda).
\end{equation*}
Looking at the definition for $\hat{A}^\noneq(\alpha,\beta,\lambda) = \avgph{e + \hat{u}/\beta} = \avgph{(\alpha_i\phi_i - f(\alpha, \beta, \lambda)}/\beta = (\alpha_i \hat{\Phi}_i - f(\alpha, \beta, \lambda))/\beta$, it's clear that $\hat{A}^\noneq(\Phi,\beta,\lambda) = f^*(\Phi,\beta,\lambda)/\beta$.  
Since $\hat{A}^\noneq(\alpha,\beta,\lambda)\neq f(\alpha,\beta,\lambda)$, the non-equilibrium free energies driving the dynamics for $\alpha$ and $\hat{\Phi}$, that is  $\hat{A}^\noneq(\alpha)$ and $\hat{A}^\noneq(\Phi)$, are not related through a Legendre transform as is typically the case when one changes between dual variables in classical thermodynamics. 

\section{An information theoretic view}
Before moving on to the example implementations of the STIV framework, we highlight an  information theoretic perspective on the dynamical equations for the internal variables based on minimizing the Kullback–Leibler (KL) divergence between the approximate and true probability density of states. Following Bronstein and Koeppl\citep{bronstein2018variational}, we assume that the true probability density of states if governed by the equation $\pd_t p = \fL p$, and try to minimize the KL-divergence 
\begin{equation*}
\mathbb{D}_{KL}(\alpha) \equiv \mathbb{D}_{KL}[\hat{p}||p]=  \int\hat{p}\log(\frac{\hat{p}}{p})\diff x.
\end{equation*}
Differentiating in $\alpha_i$ gives 
\begin{equation*}
0 = \int\log(\frac{\hat{p}}{p}) \pd_{\alpha_i}\hat{p}\, \diff x.
\end{equation*}
We would like this to be true for all time, and hence we differentiate in time to get 
\begin{align*}
0&=\int \left\{ \log(\frac{\hat{p}}{p}) \pd_{\alpha_i}\pd_{\alpha_j} \hat{p} + \hat{p}\,\pd_{\alpha_i}\hat{u}\pd_{\alpha_j }\hat{u}\right\}\dot{\alpha}_j\diff x \\
&+\int\left\{\log(\frac{\hat{p}}{p})\pd_{\alpha_i}\pd_\lambda \hat{p} + \hat{p}\pd_{\alpha_i}\hat{u}\pd_\lambda \hat{u}\right\}\dot{\lambda}- \int \pd_{\alpha_i}\hat{u}\frac{\fL p}{p} \hat{p} \diff x
\end{align*}
or in terms of expectations with respect to $\hat{p}$, 
\begin{align*}
\avgph{\pd_{\alpha_i}\hat{u}\frac{\fL p}{p}} &= \avgph{\log(\frac{\hat{p}}{p})\frac{\pd_{\alpha_i}\pd_{\alpha_j}\hat{p}}{\hat{p}}}\dot{\alpha}_j + \mathbb{C}_{ij}\dot{\alpha}_j \\
&+\avgph{\log(\frac{\hat{p}}{p})\frac{\pd_{\alpha_i}\pd_{\lambda}\hat{p}}{\hat{p}}}\dot{\lambda} + \avgph{\pd_{\alpha_i}\hat{u}\pd_\lambda \hat{u}}\dot{\lambda} 
\end{align*}
Since this equation contains $p$, it is unlikely that one could simplify it without significant approximation. Moreover, the best approximation one has for $p$ at any given time within our framework is $\hat{p}$. If we make this substitution in the previous equation, we achieve
\begin{equation*}
\mathbb{C}_{ij}\dot{\alpha}_j  + \avgph{\pd_{\alpha_i}\hat{u}\,\pd_\lambda\hat{u}}\dot{\lambda}= \avgph{\fL^\dagger \pd_{\alpha_i}\hat{u}}
\end{equation*}
which is exactly the original dynamical equations given by the variational method of Eyink. 
Thus, these dynamics minimize the KL-divergence between the true probability density of states and the approximate probability density of states at each instance in time.

\section{Examples}
As a demonstration of the enhanced flexibility of the modified STIV framework presented above, we compare the results of STIV using an exponential family and quasi-equilibrium approximation to the true probability density of states for two example systems. The first example is that of a colloidal protein diffusing along a string of DNA \citep{kim2013probing,singh2018elasticity}. This system is approximately described by a single particle diffusing in a one dimensional potential consisting of the product of a sinusoidal function and a decaying exponential. Mathematically, 
\begin{equation}\label{eqn:potentialSinExp}
    u(x) = c_0 \exp(-c_1 x) \sin(c_2x + c_3)  + \text{bounding box}.
\end{equation}
The parameters $c_0,\ldots,c_3$ were chosen to fit the potential shown in Fig.~6 of \citep{singh2018elasticity}, and the bounding box is specified to hold the particle within 60 base pairs of deepest minima (the values of the parameters and the bounding box are detailed in Appendix D).
This example allows us to study the approximation error during a relaxation towards equilibrium. The second example includes time dependent driving. A colloidal particle is assumed to diffuse in a sinusoidal landscape while coupled to an external driving protocol via a harmonic potential, a prototypical model of an optical trap. The external driving moves in the positive $x$ axis with constant velocity $v$. Mathematically,
\begin{equation}\label{eqn:potentialExtDrive}
    e(x,\lambda_t) = -\cos(\frac{2\pi x}{l}) + \frac{k}{2}(x - \lambda_t)^2,
\end{equation}
with $\lambda(t)=vt$ (see Appendix D for parameter details). Here, we can evaluate the accumulation of error in time since this system never reaches a steady state. 
\par
For these examples, we compare the approximation accuracy of the quasi-equilibrium approximate density as in Eq.\ \ref{eqn:QuasiEQ}, and a traditional exponential family without the energy as in Eq.\ \ref{eqn:ExpFamily}.  Since both examples are one dimensional, we chose sinusoidal Fourier expansion functions as the observables $\phi_n = \sin(n\pi (x - x_0) /L)$ where $x_0$ and $L$ are chosen to fit the bounding box in the diffusing protein example or the spacial extent of the simulation in the external driving example. Due to the computational complexity of solving the differential equation $\fL^\dagger_{\hat{u}} \chi_i = \pd_{\alpha_i}\hat{u}$ to find $\chi_i$ at each step, we choose not to implement the operator method. It is likely that this method will only become feasible for specific choices of observables for which the differential equation can be inverted exactly (as happens to be the case for a univariate Gaussian).
\par
Figure \ref{fig:2} shows results for the colloidal protein example, and Fig.~\ref{fig:3} shows those of the time dependent driving example. Three different implementations of the STIV framework are highlighted throughout. The first, labeled ``$\hat{p} \propto \exp(-\beta e + \alpha \phi)$'' and colored green, uses the quasi-equilibrium probability density of states and the dynamics obtained in Eq. \ref{eqn:dynQE}. The second, labeled ``$\hat{p}\propto \exp(\alpha \phi)$ Moment Closure'' and colored yellow, uses just the exponential family approximation and the dynamics obtained from the variational method of Eyink, Eq.~\ref{eqn:dynNoE}. Finally, the method labeled ``$\hat{p}\propto \exp(\alpha\phi)$ Gradient Flow'' and colored pink also uses the exponential family approximation but the second term on the right hand side of Eq.~\ref{eqn:dynNoE} (which is expected to vanish as the number of observables increases) is set to zero to obtain a gradient flow. The convergence of each method to the ground truth probability density of states is shown in panel A. The metric used is the time averaged total variation distance between the ground truth and the approximation
\begin{equation*}
    d(p,\hat{p}) = \frac1{2T}\int_0^T\int_{\mathbb{R}} |p(x,t) - \hat{p}(x,\alpha(t))|\,\diff x\diff t.
\end{equation*}
Although previously we have made use of the KL divergence to interpret the choice of dynamical equations for the internal variables, here we use the total variation distance to measure the error in the approximation as the numerical computation of the KL divergence tended to be poorly behaved. Fortunately, the total variation distance is bounded above by the KL divergence by \citep{pinsker1964information,bretagnolle1979estimation,canonne2022short}
\begin{equation*}
    d(p,q)\leq \text{min}\left(\sqrt{\frac1{2}\mathbb{D}_{\text{KL}}(p||q)},\sqrt{1 - \exp(-\mathbb{D}_{\text{KL}}(p||q))}\right).
\end{equation*}
In panel B of Fig.~\ref{fig:2}, we show the potential energy governing the system, along with a cartoon of the model setup. In the remaining panels C and D, snapshots of the densities at an intermediate time for two different numbers of observables are shown. In these frames, the ground truth probability density of states is shown in dark blue. At this time point, the probability density of states has not fully relaxed from the initial Gaussian distribution, and is still far from equilibrium. All methods converge to the ground truth solution, however, the quasi-equilibrium distribution outperforms the two exponential family models by an order of magnitude for 32 and 64 observables. As is expected, the interaction energy is a good observable and provides important information about the probability density of states, particularly in a problem concerning relaxation to the equilibrium state. Interestingly, the exponential family with gradient flow dynamics outperforms the moment-closure dynamics for small numbers of internal variables. As expected, this gap decreases for 64 internal variables once the approximate density has become highly flexible. Computationally, each of the three models require about the same run time (on a standard laptop computer), and run faster than computing the ground truth pde for few observables (4 to 32), but become slower for large numbers of internal variables (64) due to the need to solve linear systems involving the covariance matrix $\mathbb{C}_{ij}$. 
\par 
Figure \ref{fig:3} shows the results for the external driving example. Panel A, again, shows the total variation distance from the ground truth for each model, but now as a function of time over the whole course of the simulation, and for the approximations with 16 and 64 internal variables. After an initial increase, the error of each of the methods is relatively stable in time. This mirrors the observation made in previous implementations of the STIV framework, where the approximation seemed to recover despite passing through regions in time where the approximation was quite poor \citep{leadbetter2023}. Panel B shows the sinusoidal potential in blue and the external driving potential due to an optical trap in pink, along with a cartoon of the model setup. The remaining two panels (C and D) compare the approximate densities with 64 internal variables to the ground truth at two representative points in time. Although not shown here, it is worth noting that when fewer than 64 internal variables are used, including the energy greatly improves the approximation in this example. The external driving is slow enough that the density remains close to equilibrium at all times, and hence the energy is a very good observable.  

\begin{figure}
    \centering
    \includegraphics[width=1.0\linewidth]{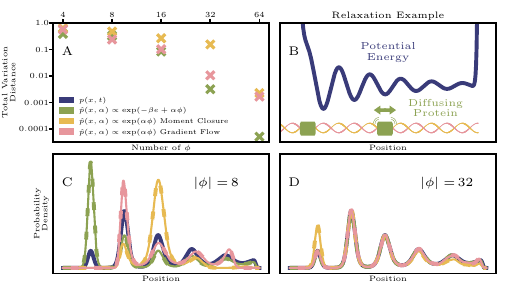}
    \caption{The convergence of STIV to the true probability density of states in a model of protein diffusion on DNA. Panel A shows the time averaged total variation distance between each STIV model (the quasi-equilibrium approximation in green, the exponential family approximation without the energy in yellow, and the exponential family approximation with modified dynamics to produce a gradient flow in pink) and the true probability density of states (obtained via a traditional solver) as a function of the number of internal variables. Panel B depicts the energy landscape. Panels C and D show the approximate probability density of states for each model for 8 and 32 internal variables respectively and the true probability density of states (in dark blue) at the same fixed time point (midway between the initial time and fully relaxed in equilibrium). }
    \label{fig:2}
\end{figure}
\begin{figure}
    \centering
    \includegraphics[width=1.0\linewidth]{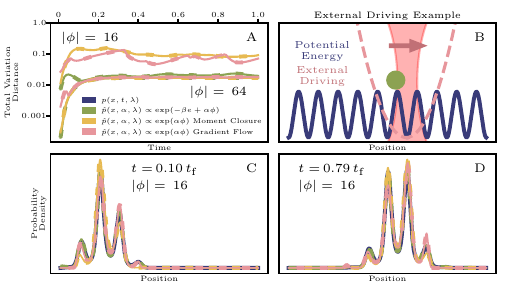}
    \caption{Investigation of the error accumulation of the STIV models for a single particle in a sinusoidal potential driven by a linear external force. Panel A shows the total variation distance between each model (same colors as Fig. \ref{fig:2}) and the true probability density of states when 64 internal variables are used. Panel B depicts the sinusoidal landscape and the quadratic potential associated with the external driving from an optical trap (at $t = .79\,t_{\text{final}}$). Panels C and D show the approximate probability density of states for each model using 64 internal variables and the true probability density of states (in dark blue) for two different points in time. ($t_{\text{final}} = 10$)}
    \label{fig:3}
\end{figure}

\section{Conclusion}
We have shown that given a microscopic particle system governed by overdamped Langevin dynamics it is always possible to construct a thermodynamically consistent macroscopic (ensemble averaged) model with a gradient flow evolution for the internal variables. One strategy, which we've called the operator method, applicable for an arbitrary choice of approximate probability density of states, is presented but not pursued due to computational challenges of a general implementation. Alternative methods, restricted to quasi-equilibrium or exponential family approximations, can be implemented relatively simply. These latter models can either be formulated in terms of averaged observables or their dual internal variables, and can approximate  continuous probability distributions with arbitrary accuracy. 
\par
Though we have shown the flexibility of this framework by considering Fourier basis functions as observables, the true utility of internal variable models comes from the intelligent choice of a few highly descriptive observables. In such cases, one can expect the STIV framework to produce accurate predictions with relatively little computational cost. Thus, a key task for continuing work is that of finding descriptive observables for a given microscopic system (see \citep{liu2023learning,qiu2025bridging} for emerging efforts in this direction).  

\section{Author contributions}
All authors have accpeted responsibility for the entire content of the manuscript and approved its submission.
\paragraph*{CRediT author statement}
\textbf{Travis Leadbetter:} Conceptualization, Methodology, Software, Formal Analysis, Writing - Original Draft, Writing - Review \& Editing. 
\textbf{Prashant K. Purohit:} Conceptualization, Writing - Review \& Editing, Supervision.
\textbf{Celia Reina:} Conceptualization, Methodology, Writing - Review \& Editing, Supervision.

\section{Conflicts of interest}
There are no conflicts of interest to declare.
\section{Data availability}
Source code used to generate the data and figures displayed in this work is freely available via the GitHub repository celiareina/STIV-GradFlow. 

\section{Acknowledgements}
T.\ L.\ acknowledges funding support from the University of Pennsylvania Applied Mathematics and Computational Science graduate group. C.~R.~gratefully acknowledges support from NSF CAREER Award (CMMI-2047506). P.\ K.\ P.\ acknowledges partial support from NSF Grant DMR 2212162.
} 

\clearpage
\renewcommand*{\thesection}{\Alph{section}}
\setcounter{section}{0}

\textbf{\LARGE Appendices} \\
\vspace{12 pt} 
\\ 
\section{The variational method of Eyink with external driving}\label{sec:SI_modEyink}
In the original derivation of the variational method of Eyink \citep{eyink1996}, when the true solution to the Fokker-Planck equation $\pd_t p(x,t) = \fL p(x,t)$ is approximated by a parameterized density of states $\hat{p}(x,\alpha)$, it was assumed that all of the time dependence of $\hat{p}$ is captured by the parameters $\alpha(t)$. However, if we wish to implement a quasi-equilibrium approximation $\hat{p} = \exp(-\beta e + \alpha_i(t) \phi_i(x) - f(\alpha,\beta))$ and the energy depends on a time-dependent external driving protocol $e = e(x,\lambda(t))$, we must modify the variational method since the time dependence assumption has been broken. 
\par 
Assume for fixed external driving protocol $\lambda(t)$, we approximate the density of states via $\hat{p}(x,\alpha,\lambda)$ (here $\beta$ is assumed to be fixed), and approximate test function $\hat{\psi}(x,\alpha)$. The variational action is $S(\alpha) = \int_0^\infty  \int_X \hat{\psi}\left(\pd_t  - \fL\right)\hat{p}\, \diff x \diff t$. Stationary points of the action are given by 
\begin{align*}
    0 &= \pd_{\alpha_i} S(\alpha) \\ 
     &= \int_0^\infty \int_X \left\{ \pd_{\alpha_i}\hat{\psi}\pd_t \hat{p} - \pd_{\alpha_i}\hat{p}\pd_t \hat{\psi}  \right\}\, \diff x \diff t - \pd_{\alpha_i} \int_0^\infty \int_X \hat{\psi}\fL \hat{p}\, \diff x \diff t \\ 
     &= \int_0^\infty \int_X \left\{ \pd_{\alpha_i}\hat{\psi}\pd_{\alpha_j}\hat{p}\dot{\alpha}_j + \pd_{\alpha_i}\hat{\psi}\pd_\lambda \hat{p} \dot{\lambda} - \pd_{\alpha_i}\hat{p} \pd_{\alpha_j} \hat{\psi} \dot{\alpha}_j \right\} \\
     &\qquad - \int_0^\infty \pd_{\alpha_i}\avgph{\fL^\dagger \hat{\psi}} \diff t \\
     &= \int_0^\infty \left\{ \avgph{[\pd \hat{\psi},\, \pd \hat{u}]_{ij}}\dot{\alpha}_j + \avgph{\pd_{\alpha_i}\hat{\psi}\pd_{\lambda}\hat{u}}\dot{\lambda} - \pd_{\alpha_i}\avgph{\fL^\dagger \hat{\psi}} \right\} \, \diff t.
\end{align*}
The proper dynamics are recovered by setting the integrand equal to zero. 
\par
We now assume the ``linear ansatz'' for the test function by introducing the paired set of parameters $(\alpha,\gamma)$  and write $\hat{\psi} = 1 + \gamma_i \pd_{\alpha_i}\hat{u}$. Plugging this in to the dynamical equation above leads to two coupled equations 
\begin{align*}
    \gamma_k \pd_{\alpha_i}\avgph{\fL^\dagger \pd_{\alpha_k}\hat{u}} &= \gamma_k\avgph{\pd_{\alpha_i}\pd_{\alpha_k}\hat{u}\pd_{\alpha_j}\hat{u} - \pd_{\alpha_j}\pd_{\alpha_k}\hat{u}\pd_{\alpha_i}\hat{u}}\dot{\alpha}_j \\
    &\qquad - \avgph{\pd_{\alpha_i}\hat{u}\pd_{\alpha_j}\hat{u}}\dot{\gamma}_j + \gamma_k \avgph{\pd_{\alpha_i}\pd_{\alpha_k}\hat{u}\pd_\lambda \hat{u}}\dot{\lambda} \\
    \avgph{\fL^\dagger \pd_{\alpha_i}\hat{u}} &= \avgph{\pd_{\alpha_i}\hat{u}\pd_{\alpha_j}\hat{u}}\dot{\alpha}_j + \avgph{\pd_{\alpha_i}\hat{u}\pd_\lambda \hat{u}}\dot{\lambda}.
\end{align*}
The first equation is trivially solved by $\gamma_i \equiv 0$, whereas the second equation is independent of $\gamma_i$ and gives the resulting dynamics we need. 
\section{Approximate thermodynamic quantities for the quasi-equilibrium approximation of a system with external driving}
Since the inclusion of the energy in the approximate probability density of states breaks the assumption that $\hat{p}$ be independent of $\lambda$, the form of the approximate thermodynamic quantities must change as a result. Recall that the quasi-equilibrium approximation takes the form $\hat{p} = \exp(-\beta e(x,\lambda) + \alpha_i \phi_i(x) - f(\alpha,\beta,\lambda))$ so that $\pd_{\alpha_i}\hat{u} = \phi_i - \hat{\Phi}_i$, $\hat{a}^\noneq = e + \hat{u}/\beta = (\alpha_i \phi_i - f)/\beta$ and, following the same steps as in the main text, $\avgph{\fL^\dagger\pd_{\alpha_i}\hat{u}} = - \frac1{\eta\beta}\mathbb{M}_{ij}\alpha_j$ where $\mathbb{M}_{ij} = \avgph{\pd_x \phi_i \pd_x \phi_j}$.  Plugging this into the dynamical equations derived from the previous section leads to
\begin{align*}
    \mathbb{C}_{ij}\dot{\alpha}_j + \avgph{(\phi_i - \hat{\Phi}_i)\pd_{\lambda}\hat{u}}\dot{\lambda} = -\frac1{\eta\beta}\mathbb{M}_{ij}\alpha_j
\end{align*}
where $\mathbb{C}_{ij} = \avgph{(\phi_i - \hat{\Phi}_i)(\phi_j - \hat{\Phi}_j)}$. 
The approximate external force is defined as $\hat{F}^\text{ex} \equiv \avgph{\pd_\lambda e}$ which is no longer simply $\pd_\lambda \hat{A}^\text{neq}$. Now 
\begin{align*}
    \pd_{\lambda} \hat{A}^\noneq &= \avgph{\pd_\lambda e} + \avgph{(e + \hat{u}/\beta)\pd_{\lambda}\hat{u}} \\
    &= \hat{F}^\text{ex} + \frac1{\beta}\avgph{\alpha_i\phi_i \pd_{\lambda}\hat{u}}\\
    &= \hat{F}^\text{ex} + \frac1{\beta}\pd_\lambda (\alpha_i \hat{\Phi}_i)
\end{align*}
which gives the first part of Eq.\ \ref{eqn:FEx} of the main text. Since $\hat{A}^\noneq = (\alpha_i\hat{\Phi}_i - f)/\beta$, $\hat{A}^\noneq - \alpha_i\Phi_i/\beta = -f/\beta$ and we recover the second part. 
\par
To identify the form of the entropy production, we return to its definition from stochastic thermodynamics and its approximation via STIV. The incremental entropy production is the difference in the incremental work minus the incremental free energy, $T\diff s^\text{tot} = \diff w - \diff a^\noneq$. Approximating these terms at the ensemble average level leads to the required relation $T\frac{\diff}{\diff t} \hat{S}^\text{tot} = \frac{\diff}{\diff t} \hat{W} - \frac{\diff}{\diff t}\hat{A}^\noneq$. The work rate is identified at $\frac{\diff}{\diff t}\hat{W} = \hat{F}^\text{ex}\dot{\lambda}$ and the rate of change in the non-equilibrium free energy writes $\frac{\diff}{\diff t}\hat{A}^\noneq = \pd_\lambda \hat{A}^\noneq \dot{\lambda} + \pd_\alpha \hat{A}^\noneq \dot{\alpha}$. Inserting $\pd_\lambda \hat{A}^\noneq = \hat{F}^\text{ex} + \alpha_i\avgph{(\phi_i - \hat{\Phi}_i)\pd_\lambda \hat{u}}/\beta$ into the equation for $T\frac{\diff}{\diff t} \hat{S}^\text{tot}$ leaves (note $\pd_\lambda \hat{u}$ has mean zero due to normalization)
\begin{align*}
    T\frac{\diff}{\diff t} \hat{S}^\text{tot} = -\dot{\alpha}\pd_\alpha \hat{A}^\noneq - \frac{\alpha_i}{\beta}\avgph{ (\phi_i - \hat{\Phi}_i)\pd_\lambda \hat{u}}\dot{\lambda}.
\end{align*}
Since $\pd_{\alpha_i}\hat{A}^\noneq = \mathbb{C}_{ij}\alpha_j/\beta$, we can multiply the whole dynamical equation by $-\alpha_i/\beta$ to see 
\begin{align*}
    T\frac{\diff}{\diff t} \hat{S}^\text{tot} = \frac{\alpha_i\mathbb{M}_{ij}\alpha_j}{\eta\beta^2} 
\end{align*}
giving Eq.\ \ref{eqn:EPQEq} in the main text. 

\section{Proof that exponential families and quasi-equilibrium distributions can approximate any continuous probability density on $\mathbb{R}^n$ in total variation distance} \label{sec:SI_expFam}
Let $p$ be a continuous probability density on $\mathbb{R}^n$ and let $u = \log(p)$. Since $p$ is integrable, for any fixed $0 < \epsilon < 1$, we can find a compact subset $K \subset \mathbb{R}^n$ such that $K^c = \mathbb{R}^n \setminus K$  obeys $\int_{K^c} p(x) \diff x < \epsilon$/2. Let $K' = K\cap\{p\geq \frac{\epsilon}{2\text{Vol}(K)}\}$. $K'$ is compact since $p$ is continuous, and we know 
\begin{align*}
    \int_{K'^c} p(x)\diff x &= \int_{K^c}p(x)\diff x + \int_{K \cap \{p < \epsilon/2\text{Vol}(K)\}} p(x)\diff x \\
    &< \epsilon/2 + \epsilon/2 = \epsilon.
\end{align*}
$u$ is bounded on $K'$, so by the Stone-Weierstrass theorem, we can find a polynomial $q(x)$ such that $||{u - q}||_{\fL^\infty(K')} = \sup_{x \in K'}|u(x) - q(x)| < \epsilon$. As $e^{q(x)} = e^{u(x)}e^{q(x) - u(x)} \leq e^{u(x)}e^\epsilon$, $Z_q \equiv \int_{K'} e^{q(x)}\diff x < \infty$ and we can define $\hat{p} = \Indic_{x\in K'}\frac{e^{q(x)}}{Z_q}$. In fact, we have $e^{-\epsilon}(1 - \epsilon) \leq Z_q \leq e^\epsilon$. Now 
\begin{align*}
    d_{\text{TV}}(p,\hat{p}) &= \frac1{2}\int|p(x) - \hat{p}(x)|\diff x \\
    &= \frac1{2}\int_{K'} |p(x) - \hat{p}(x)|\diff x + \frac1{2}\int_{K'^c}p(x)\diff x \\
    &\leq \frac1{2}\int_{K'} e^{u}\left|1 - \frac{e^{q - u}}{Z_q}\right|\diff x + \frac{\epsilon}{2} \\
    &\leq \frac1{2}||1 - \frac{e^{q - u}}{Z_q}||_{\fL_\infty(K')}\int_{K'} p\,\diff x + \frac{\epsilon}{2} \\
&\leq \frac1{2}\left(e^{2\epsilon} - 1\right) + \epsilon \rightarrow 0 
\end{align*}
as $\epsilon \rightarrow 0$ since $||1 - \frac{e^{q - u}}{Z_q}||_{\fL^\infty(K')}\leq \text{max}\left(\frac{e^{2\epsilon}}{1 - \epsilon}-1,1 - e^{-2\epsilon}\right) = \frac{e^{2\epsilon}}{1 - \epsilon}-1 $. 
\par 
An analogous proof shows that the same is true for quasi-equilibrium distributions. Assuming that the energy is continuous as well, we simply choose $K'$ as above,  $q$ such that $||u + \beta e - q||_{\fL^\infty(K')} < \epsilon$, define $Z_q = \int_{K'} e^{-\beta e + q} \diff x$. and let $\hat{p} = \Indic_{x \in K'} \frac{e^{-\beta e + q}}{Z_q}$. The same bounds hold for $Z_q$ so we again have 
\begin{equation*}
    d_{\text{TV}}(p,\hat{p}) \leq \frac1{2}\left(e^{2\epsilon} - 1\right) + \epsilon \rightarrow 0.
\end{equation*}

\section{Details of the numerical implementation of STIV for examples}
As discussed in the main text, the energy used to study relaxation to equilibrium comes from a model of the energy landscape of a protein diffusing on a strand of DNA. The original experimental data was published by \cite{kim2013probing}, and the particular choice of model energy landscape was taken from \cite{singh2018elasticity}. The parameters used in Eq.\ 13 are $c_0 = -0.87817$, $ c_1 = 0.04736$, $c_2 = 0.60955$, $c_3 = 1.25408$. A bounding box of the form (where $x$ has units of base pairs)
\begin{equation*}
    \text{Bounding Box}(x) = 100\,e^{-x - 40}  \mathbf{1}_{(x < -25)} + \left(\frac{(x + 10)}{30}\right)^{100}
\end{equation*}
was used to constrain the simulated proteins to $x \in [-40,20]$ (that is, the initial range of the model energy landscape in \cite{singh2018elasticity}) and to insure that the minima at $x \approx -30$ was the left most minima. This range is translated to $[0,60]$ in the video in Appendix E.  Since the energy has units of $k_BT$, we used $\beta = 1$. We chose $\eta = .83$ which is artificially large to accelerate all simulations.  
\par
For the external driving example, we chose parameters $l = 1.0$, $k = 1.6287$, $\beta = 1.$, and $\eta = .8298$. The external protocol was chosen as $\lambda(t) = .5 t$. 
\par
All models (both STIV and the ground truth) were implemented on a discrete grid of 2000 equally spaced points (between $[-39,21.1]$ for the relaxation example and $[-2,8]$ for the external driving example). The ground truth was obtained from discretizing the true pde, $\pd_t p = \fL p$, and solving the resulting ode using the Dormand-Prince adaptive Runge-Kutta scheme with variable step size to ensure the relative error remained less than $0.001$. For the STIV models, the energy and observables were computed on the grid points $e_i = e(x_i) \in \mathbb{R}^{2000}$ and $\phi_{ij} = \phi_j(x_i) \in \mathbb{R}^{2000 \times d}$ where $d$ is the number of internal variables used. $\hat{p}_i = \hat{p}(x_i,\alpha)$ was then approximated as
\begin{equation*}
    \hat{p}_i = \frac1{\Delta x}\exp(-\beta e_i + \phi_{ij}\alpha_j - f(\alpha,\beta))
\end{equation*}
with 
\begin{equation*}
    f(\alpha,\beta) = \log(\sum_{i=1}^{2000} \exp(-\beta e_i + \phi_{ij}\alpha_j)).
\end{equation*}
Any expectations in the dynamical equations were then approximated using the discrete approximate density of states 
\begin{equation*}
    \avgph{g} \approx \sum_{i=1}^{2000}g(x_i)\hat{p}_i \Delta x.
\end{equation*}
The exponential family models which did not include the energy were obtained the same way but with $\beta$ fixed to zero. 
\par 
Some numerical difficulties arose while inverting the covariance matrix $\mathbb{C}_{ij}$. Thus, $\mathbb{C}^{-1}_{ij}$ was further approximated using the pseudo-inverse $\mathbb{C}_{ij}^+$ truncating singular values (of $\mathbb{C}_{ij}$) falling below $10^{-8}$. Various truncation levels ($10^{-2}$ through $10^{-15}$) were tested to ensure this choice did not impact numerical accuracy.
\clearpage
\section{Full video of examples}
This video can be found with the supplemental information, or is available upon reasonable request. 
\par
This video shows a comparison between the true probability density of states in dark blue and the approximate densities using the quasi-equilibrium method (green), the exponential family approximation with dynamics given by the variational method (equivalent to moment closure, yellow), and the exponential family approximation with dynamics given by a gradient flow (salmon). The first part of the video shows, sequentially, the approximation for the model of proteins diffusing on DNA using 4, 8, 16, 32, and 64 internal variables (the legend $|\phi| = N$ in each part shows the number of internal variables/observables). The second half of the video shows the approximation for the externally driven particle in a periodic landscape for 4, 8, 16, 32, and 64 internal variables.

\printbibliography

\end{document}